\documentclass[aps,pra,notitlepage,twocolumn,superscriptaddress]{revtex4-1} 
\usepackage{amsmath,epsfig,color,amssymb}
\usepackage{graphicx}
\usepackage{graphicx}
\usepackage{dcolumn}
\usepackage{bm}
\usepackage{graphicx}
\usepackage{amsmath, amsfonts, amssymb,mathrsfs}
\usepackage{pstricks}
\usepackage{amsxtra}
\usepackage{amsthm}
\usepackage{natbib}

\def\be{\begin{equation}}      
\def\ee{\end{equation}}
\def\beu{\begin{equation*}}   
\def\eeu{\end{equation*}}

\providecommand{\abs}[1]{\left\lvert#1\right\rvert}   
\providecommand{\ket}[1]{\left|#1\right\rangle}

\providecommand{\bra}[1]{\left\langle#1\right|}

\definecolor{new}{rgb}{0,0,1}
\definecolor{old}{rgb}{1,0,0}

\begin{document}
\title{Protocol for a resonantly-driven three-qubit Toffoli gate with silicon spin qubits}
\author{M. J. Gullans}
\author{J. R. Petta} 
\affiliation{Department of Physics, Princeton University, Princeton, New Jersey 08544, USA}

\begin{abstract}
The three-qubit Toffoli gate plays an important role in quantum error correction and complex quantum algorithms such as Shor's factoring algorithm, motivating the search for efficient implementations of this gate.  Here we introduce a Toffoli gate suitable for exchange-coupled electron spin qubits in silicon quantum dot arrays.  Our protocol is a natural extension of a previously demonstrated resonantly driven CNOT gate for silicon spin qubits.  It is based on a single exchange pulse combined with a resonant microwave drive, with an operation time on the order of 100 ns and fidelity exceeding 99\%.  We analyze the impact of calibration errors and $1/f$ noise on the gate fidelity and compare the gate performance to Toffoli gates synthesized from two-qubit gates.  Our approach is readily generalized to other controlled three-qubit gates such as the Deutsch and Fredkin gates.  
\end{abstract}
\maketitle

\section{Introduction}
Silicon quantum dots provide a clear path towards scalable quantum information processing with spin qubits \cite{Loss98,Hanson07,Zwanenburg13}.  At low temperatures, electron spins in isotopically enriched silicon can have coherence times exceeding a second \cite{Tyryshkin12}.  Single qubit gate fidelities can exceed 99.9\% \cite{Yoneda18,Yang18} and there have been several recent demonstrations of fast two-qubit gates \cite{Veldhorst15,Zajac18,Watson18,Xue18} with fidelities reaching  98\% \cite{Huang18}.  Concurrent with these developments have been improvements in scaling up to large arrays of individually controllable quantum dots coupled through nearest-neighbor exchange interactions \cite{Zajac16,Hensgens17,Mortemousque18,Mukhopadhyay18,Mills18,Volk19}, with prospects for long-range spin-spin coupling using superconducting resonators \cite{Mi18,Landig18,Samkharadze18}.   Future milestones will include the demonstration of basic  quantum algorithms in systems of three or more silicon spin qubits.  

A key challenge in implementing large scale quantum algorithms is developing efficient gate compilation strategies to improve performance and reduce the overhead in implementing fault tolerant gates \cite{Svore06,Jones12,Haner18}.  The three-qubit Toffoli gate (controlled$-$CNOT gate) is a universal gate for reversible classical computation. It also plays an important role in quantum error correction and Shor's factoring algorithm \cite{NielsenChuang,Cory98,Vandersypen01}.  It is therefore desirable to realize fast, high-fidelity implementations of the Toffoli gate with spins in silicon. Successful demonstrations of the Toffoli gate have been achieved in a number of quantum information platforms \cite{Cory98,Vandersypen01,Monz09,Lanyon09,Fedorov12,Shi18,Beterov18}. However, a detailed protocol for implementing a Toffoli gate with silicon spin qubits has not been put forward.

\begin{figure}[b]
\begin{center}
\includegraphics[width=0.49\textwidth]{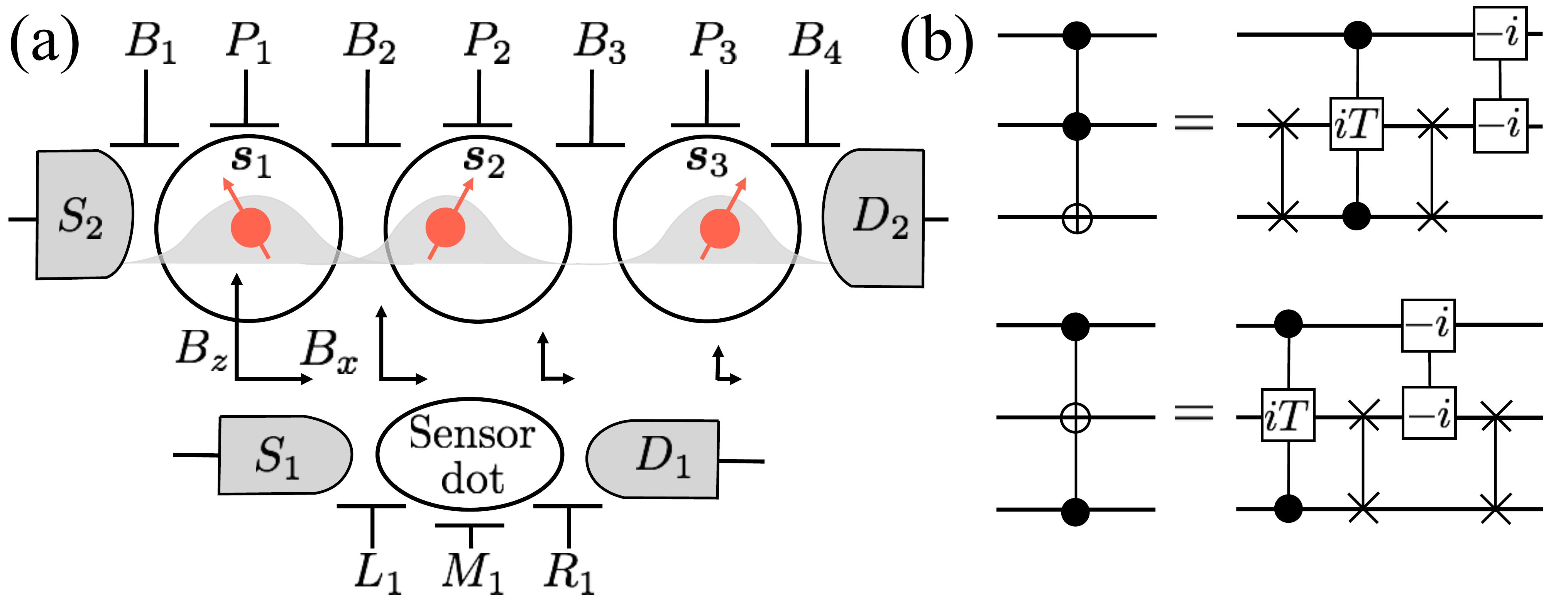}
\caption{(a) Electrically-controllable spin qubits arranged in a linear array. Locally varying magnetic fields ($B_x$ and $B_z$) allow for site-selective quantum control. Electric dipole spin resonance (EDSR) is achieved by driving the plunger gates $P_i$ with microwave fields. Nearest-neighbor exchange coupling is controlled via barrier gates $B_i$ and an exchange gate between spins 1 and 2 is depicted here.  A sensor dot allows readout of the spin state of each dot via spin-to-charge conversion. (b) Quantum circuit diagram for the implementation of the Toffoli gate with an edge (top panel) or middle (bottom panel) qubit chosen as the target. In our approach four multi-qubit gates are needed to implement a Toffoli gate: one $i$-Toffoli ($iT)$, two SWAPs, and one $-i$ C-Phase.}
\label{fig:schematic}
\end{center}
\end{figure}

Here we introduce an efficient implementation of the Toffoli gate suitable for three individually addressable exchange-coupled Loss-DiVincenzo spin qubits \cite{Loss98}.  Our implementation is a natural extension of a previously demonstrated CNOT gate for silicon spin qubits \cite{Zajac18,Russ18}. It is based on an exchange pulse combined with a microwave drive applied to a target spin whose electron spin resonance condition depends on the states of the two neighboring spins. In a linear array of triple quantum dots [see Fig.~\ref{fig:schematic}(a)], the central spin most naturally serves as the target qubit since it is exchange coupled to two nearest neighbor spins. If desired, the target qubit can be moved to one of the other two dots using SWAP gates. We present a detailed analysis of the calibration conditions for this gate and compare its performance to Toffoli gates synthesized from two-qubit gates.  Our approach can be used to realize other controlled three-qubit gates such as the Deutsch and Fredkin gates \cite{Fredkin82,Milburn89,Deutsch89}.

\begin{figure*}[tb]
\begin{center}
\includegraphics[width=0.85\textwidth]{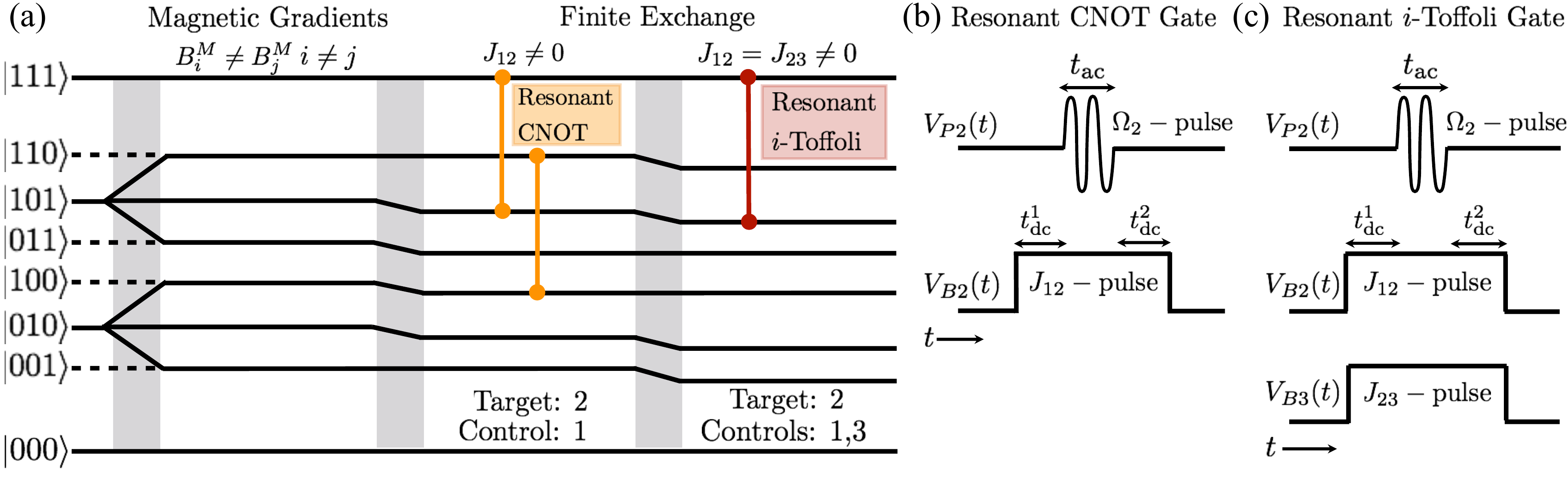}
\caption{(a) Level diagram for the 3-qubit system, where $\ket{s_1 s_2 s_3}$ specifies the 3-qubit spin state and 0/1 corresponds to spin-down/spin-up. Lines denote spectroscopically-resolved transitions used to drive the resonant CNOT and resonant $i$-Toffoli gates.  (b) Voltage control sequence to realize the resonant CNOT gate studied in Refs.~\cite{Zajac18,Russ18}.  (c) Voltage control sequence to realize a resonant $i$-Toffoli gate.  Exchange couplings $J_{12}$ and $J_{23}$ are pulsed on for qubit pairs (1,2) and (2,3).  After exchange is turned on, spin $s_2$ is flipped by a resonant EDSR pulse conditioned on the state of spins $s_1$ and $s_3$.  Additional phases on the qubits are cancelled out by carefully choosing the pulse times $t_{\rm ac}$, $t_{\rm dc}$ and the ratio of $\Omega_2/J_{i j}$, as well as by setting $J_{12} = J_{23}$.}
\label{fig:gate}
\end{center}
\end{figure*}

The paper is organized as follows: In Sec.~\ref{sec:overview} we present an overview of the basic spin-qubit architecture we consider and the physics underlying our implementation of the Toffoli gate. We then present a detailed discussion  of the calibration procedures for the gate and compute the average gate fidelities in the presence of calibration errors in Sec.~\ref{sec:tuning}.  In Sec.~\ref{sec:noise} we analyze the performance of the gate in the presence of $1/f$ charge noise. In Sec.~\ref{sec:syn}, we contrast the performance of the resonantly driven Toffoli gate with Toffoli gates synthesized from two qubit gates \cite{Barenco95}. We present our outlook and conclusions in Sec.~\ref{sec:conc}.

\section{Toffoli Gate Implementation}
\label{sec:overview}

A schematic of the system, which is based on three exchange coupled semiconductor quantum dots, is shown in Fig.~\ref{fig:schematic}(a) \cite{Zajac16}.  A magnetic field gradient in the $z$-direction enables site selective control of each spin, while a transverse field gradient in the $x$-direction allows for single spin rotations through electric dipole spin resonance (EDSR) \cite{Rashba08}.  Plunger gates $P_{1}$ through $P_3$ are used to control the occupancy of each dot and drive EDSR.  Barrier gates $B_2$ and $B_3$ control the exchange interaction between nearest-neighbor quantum dots 1 and 2, and 2 and 3, respectively. Spin-state readout can be performed using spin-to-charge conversion with a charge sensor quantum dot \cite{Hanson07}.

The quantum circuit for a Toffoli gate is shown in Fig.~\ref{fig:schematic}(b).  Its action on the qubit basis states takes the form 
\be
\ket{a,b,c} \to  \ket{a,b,c \otimes ab}
\ee
where $a,b,c \in \{0,1\}$ and $\otimes$ is the logical exclusive$-$OR operation and $0/1$ corresponds to spin-down/spin-up.    In our approach, the Toffoli gate can be synthesized from two SWAP gates,  a native three qubit gate that we refer to as the $i$-Toffoli gate  $iT$, and the $-i$ C-Phase gate. The action of the $i$-Toffoli gate on the qubit basis states is given by 
\be
iT\ket{a,b,c} = i^{ab} \ket{a,b,c\otimes ab}.
\ee
\noindent The $-i$ C-Phase gate is needed to cancel the factor of $i$ in the $iT$ gate.  Due to the lack of next-nearest-neighbor interactions, the $i$-Toffoli is most naturally implemented with the central qubit chosen as the target.

In the absence of an EDSR drive, the low-energy Hamiltonian for the three dot system takes the form \cite{Russ18} 
\be
H = \sum_i \bm{B}_i \cdot \bm{s}_i  + \sum_{i} J_{i i+1}( \bm{s}_i \cdot \bm{s}_{i+1} -1/4),
\ee
\noindent where we have set $\hbar$ and $g^* \mu_B $ equal to one $(g^* \approx 2$ is the electron $g$-factor in Si), $\bm{B}_i$ is the local magnetic field of dot $i$, and $J_{i i+1}$ is the exchange interaction between dots $i$ and $i+1$.
The spectrum of $H$ for a three-qubit system at large magnetic fields $B_i^z = B^{\rm ext} + B_i^M$ is shown in Fig.~\ref{fig:gate}(a).  Here $B_i^M$ is the magnetic field generated by the micromagnet at site $i$ and $B^{\rm ext}$ is the external magnetic field applied along the $z$-direction. The far left side of the energy level diagram illustrates the case of a uniform magnetic field, with no
exchange and no field gradient. Here the $\ket{000}$ and $\ket{111}$ states are split off by the Zeeman energy, while the $m_s = -1/2$ manifold $\{\ket{100},\, \ket{010},\, \ket{001}\}$ and the $m_s = 1/2$ manifold $\{ \ket{110},\, \ket{101},\, \ket{011}\}$  both have a threefold degeneracy. The $z$-component of the magnetic field from the micromagnet lifts the degeneracies of each of these manifolds allowing site-selective control of each qubit. However, the field gradient on its own does not enable controlled rotations. 

Controlled rotations are made possible by the combination of field gradients and nearest neighbor exchange coupling.  When the exchange interaction is small compared to the magnetic field gradient \cite{Meunier11}, turning on $J_{12}$ shifts the states $\ket{01 s_3}$ and $\ket{10 s_3}$ down in energy and enables a resonantly-driven two-qubit CNOT gate with the control sequence shown in Fig.~\ref{fig:gate}(b) \cite{Zajac18,Russ18}.  Additionally turning on $J_{23}$ shifts the states $\ket{s_1 0 1}$ and $\ket{s_1 10}$ down in energy, leading to a spectroscopically distinct 3-spin transition $\ket{101} \to \ket{111}$.  Applying a $\pi$-pulse resonant with this transition directly leads to the $i$-Toffoli gate, where the second spin is flipped only if the two other qubits are spin up. For a spin-1/2 system, a 2$\pi$ rotation brings the state back to itself up to a minus sign or phase of $\pi$; therefore, a $\pi$ pulse in the $\{\ket{101},\ket{111}\}$ subspace  naturally leads to the factor of $i$ inherent to the $i$-Toffoli gate.

\section{Toffoli Gate Tune-up}
\label{sec:tuning}

To physically realize the $i$-Toffoli gate we use the control sequence shown in Fig.~\ref{fig:gate}(c).  Two square wave voltage pulses are applied to the barrier gates ($B_2$ and $B_3$) to turn on  nearest-neighbor exchange interactions ($J_{12}$ and $J_{23}$). Driving plunger gate $P_2$ with a microwave field results in EDSR on the second qubit with Rabi frequency $\Omega_2$.  Following the approach outlined in Ref.~\cite{Russ18} for the resonant CNOT gate, we find that after appropriately controlling the drive time and Rabi frequency to minimize unwanted population transfers, this sequence of operations can be used to drive a high-fidelity $\pi$-pulse in the $\{ \ket{1 0 1}, \ket{111}\}$ subspace.  In addition to these dynamics, however, there are also phases that are accumulated on each of the eight three-qubit states.  Similar to the resonant CNOT gate, these phases can be compensated via careful calibration of pulse lengths ($t^1_{dc}$, $t^2_{dc}$, and $t_{ac}$), the Rabi frequency $\Omega_2$, and through the application of single-qubit $Z$ rotations after the controlled rotation.  The $Z$ rotations may be implemented in software. We now present a detailed analysis of these calibration conditions and the robustness of the gate to calibration errors.

In the presence of a magnetic field gradient that is large compared to exchange, and an EDSR microwave drive applied to qubit 2, the Hamiltonian in the rotating wave approximation for the three-spin system can be decomposed as a direct sum of four effective two level systems 
\begin{align}
H_{(0, 0) } &= - \frac{\Delta_1  + \Delta_3  }{2} \mathbb{I} +( \Delta_2 -\bar{J} - \delta_2) s_2^z + \Omega_2 s_2^x  , \\
H_{(0, 1) } &=  \frac{\Delta_3  - \Delta_1  }{2} \mathbb{I} +( \Delta_2 -\delta {J}- \delta_2) s_2^z +\Omega_2 s_2^x , \\
H_{(1, 0) } &=  \frac{\Delta_1  - \Delta_3  }{2} \mathbb{I} +( \Delta_2 + \delta{J}- \delta_2) s_2^z +\Omega_2 s_2^x  , \\
H_{(1, 1) } &=  \frac{\Delta_1  + \Delta_3  }{2} \mathbb{I} +( \Delta_2 + \bar{J} - \delta_2) s_2^z + \Omega_2 s_2^x  , 
\end{align}
\noindent where $H_{(s_1,s_3)}$ is the projected Hamiltonian when qubits (1,3) are in the states $(s_1,s_3)$.  The parameters in $H_{(s_1,s_3)}$ are defined as follows:  $\Delta_i = B_i^z - \omega_{0i}$ is the local magnetic field for each dot relative to the EDSR drive frequencies $\omega_{0i}$ used for single-qubit gates performed in the absence of exchange,  $\delta_2 = \omega_2 - \omega_{02}$ is the shift of the qubit 2 EDSR drive between its value in the exchange ``off'' configuration and its value in the exchange ``on'' configuration (accounting for possible shifts in the magnetic gradients as the exchange is turned on), $\bar{J} = (J_{12}+J_{23})/2$,  $\delta J = (J_{12} - J_{23})/2$, and we have taken the EDSR drive of qubit 2 to be along the $x$-axis. With these projected Hamiltonians we can write the unitary for the gate operation in Fig.~\ref{fig:gate}(c) in a frame rotating with $\omega_{02}$ as
\begin{align}
U &= U_{\rm dc}^2 U_{\rm ac} U_{\rm dc}^1, \\
U_{\rm dc}^i &= e^{-i \sum_i \Delta_i s_i^z t_{\rm dc}^i - i (J_{12} s_1^z +J_{23} s_3^z) s_2^z t_{\rm dc}^i}, \\
U_{\rm ac} & = e^{ - i \delta_2 s_2^z t_{\rm ac}} \bigoplus_{(s_1, s_3)}e^{- i H_{(s_1, s_3)} t_{\rm ac}}.
\end{align}
The $i$-Toffoli gate is most simply realized in the limit $\Omega_2 \ll \bar{J}$ with $\Delta_2 =\delta J =0$ and taking  the EDSR drive frequency $\delta_2 = \bar{J}$ on resonance with the shifted transition in the $(s_1,s_3)=(1,1)$ subspace.  In this case, we can approximately neglect the $\Omega_2$ term in $H_{(s_1,s_3)}$ for $(s_1,s_3)\ne (1,1)$, which implies that these subspaces simply undergo diagonal phase evolution.  In the $(s_1,s_3)=(1,1)$ subspace, however,
\be
e^{-i H_{1 1} t_{\rm ac}} =  e^{ -i \Omega_2 s_2^x t_{\rm ac}} = \pm i \sigma_2^x,
\ee
where the last equality holds for  $\pi$-pulse times such that $\sin (\Omega_2 t_{\rm ac}/2)  = \mp 1$.  Correcting for the phases accumulated on each 3-qubit state allows for a direct realization of the Toffoli gate.  In practice, however, the constraint $\abs{\Omega_2} \ll \bar{J}$ is too restrictive and results in slow gate times, making the gate susceptible to calibration errors and decoherence.  We now show  how to overcome this limitation by also calibrating the ratio $\abs{\Omega_2}/\bar{J}$, which allows one to drive a high-fidelity, controlled spin-rotation on qubit 2 in the minimum amount of time for simple square wave pulse profiles.

\begin{figure}[tb]
\begin{center}
\includegraphics[width=0.49\textwidth]{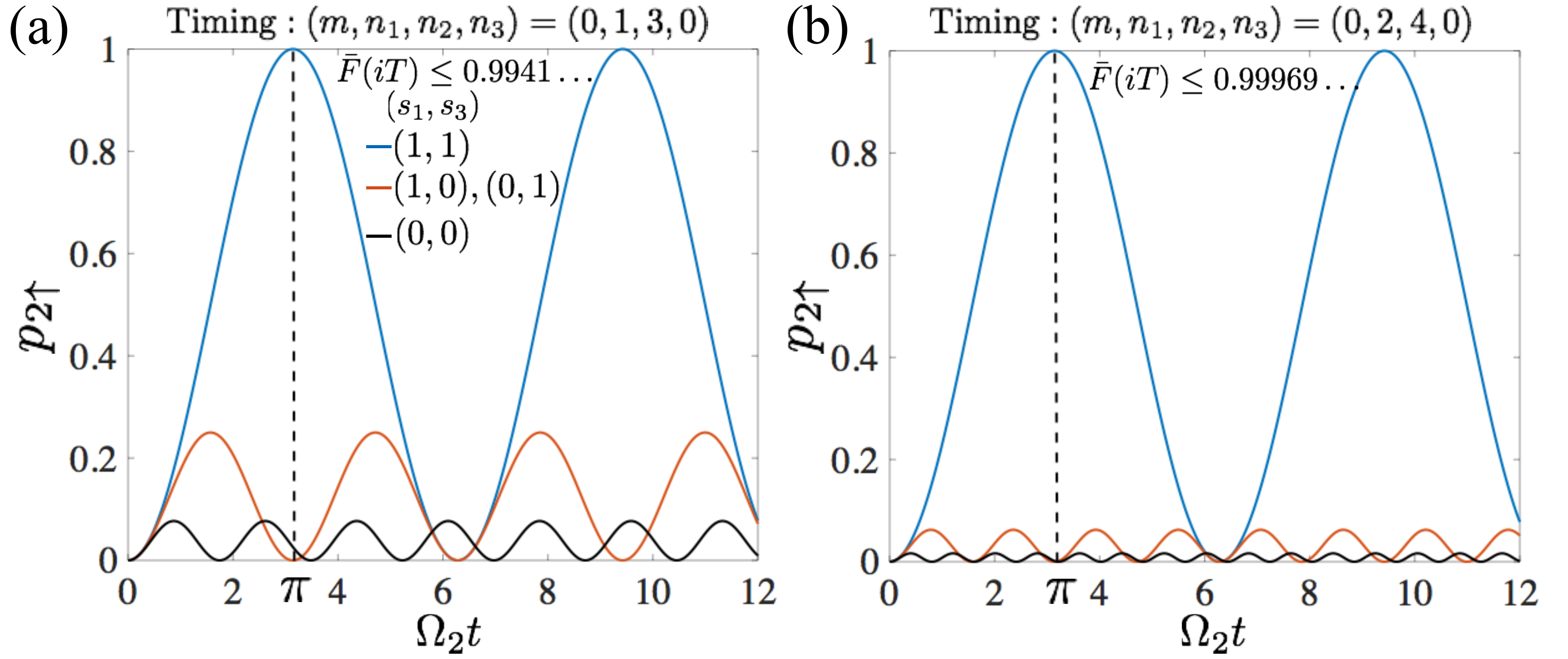}
\caption{(a) Time evolution of the spin-up probability of qubit 2, $p_{2\uparrow}$, during the application of the EDSR drive for each of the four possible spin-configurations of qubits $1$ and $3$.  Tuning to $\Omega_2 = \Omega_2(m,n_1)$ ensures that the $\pi$-pulse time for qubit 2 when both qubits $1 $ and $3$ are spin-up leads to a full 2$\pi$ rotation of qubit 2 when only qubit $1$ or $3$ are spin-down. There is only a small rotation error when qubits 1 and 3 are both spin-down.  (b) Increasing from $(n_1,n_2)=(1,3)$ to $(2,4)$  improves the fidelity from the 99\% level to the 99.9\% level.  }
\label{fig:pop}
\end{center}
\end{figure}

To ensure that the $(s_1,s_3)=(1,1)$ subspace undergoes a $\pi$ rotation, we first  impose the timing constraint $t_{\rm ac} = (2 m+1) \pi /\abs{\Omega_2}$ for integers $m \ge 0$.  We can allow $m$ to be small if we require the population dynamics in the $(s_1, s_3)\in \{(0,1),(1,0)\}$ subspaces to undergo a full $2 \pi$ rotation with respect to their precession frequencies $\Omega_{(1,0)} = \sqrt{J_{12}^2 + \Omega_2^2}$ and $\Omega_{(0,1)} = \sqrt{J_{23}^2 + \Omega_2^2}$, respectively \cite{Russ18}.  This condition can be satisfied for $J_{12} = J_{23}$ when 
\be \label{eqn:om10}
\Omega_{(0,1)} t_{\rm ac}  = (2m +1) \pi \sqrt{\bar{J}^2/\Omega_2^2 +1 } = 2 n_1 \pi ,
\ee
for an integer $n_1 >0$.   Satisfying Eq.~(\ref{eqn:om10}) requires  
\be
|\Omega_2(m,n_1)|=   \frac{(2m+1) \bar{J} }{\sqrt{4 n_1^2-(2m+1)^2} }.
\ee
 We must still ensure that there are no induced spin flips on qubit 2 in the $(s_1, s_3)=(0, 0)$ subspace.  There is no way to simultaneously satisfy a perfect $2\pi$ rotation in this subspace and the $(s_1, s_3)\in \{(0,1),(1,0)\}$ subspaces with the simple square wave pulses considered here; however, we show in Fig.~\ref{fig:pop}(a) that the off-resonant driving of the $(s_1, s_3)=(0, 0)$ subspace results in a small average gate infidelity $1-\bar{F}(iT) \sim 0.6\, \%$ for $m$ = 0 and $n_1$ = 1. Here, the average gate fidelity for unitary operators $G$ and $U$ acting on a $d$-dimensional Hilbert space  $\mathbb{C}^d$  is defined as
  \be
  \bar{F}(G) = \int d \psi \lvert \bra{\psi} G^\dag U \ket{\psi}\lvert^2, 
  \ee
\noindent where $G$ is the ideal implementation the gate, $U$ is the actual implementation, and the integral is over the Haar measure on $\mathbb{C}^d$ \cite{Nielsen02}.
 As shown in Fig.~\ref{fig:pop}(b), decreasing the Rabi frequency, by changing the timing condition to $n_1=2$, further improves the average gate infidelity to $1-\bar{F}(iT) \sim 0.03\, \%$. The maximum  fidelity increases with $n_1$ because the Rabi drive becomes weaker in this limit, resulting in less unwanted population transfer outside the target subspace.

\begin{figure}[tb]
\begin{center}
\includegraphics[width=0.49\textwidth]{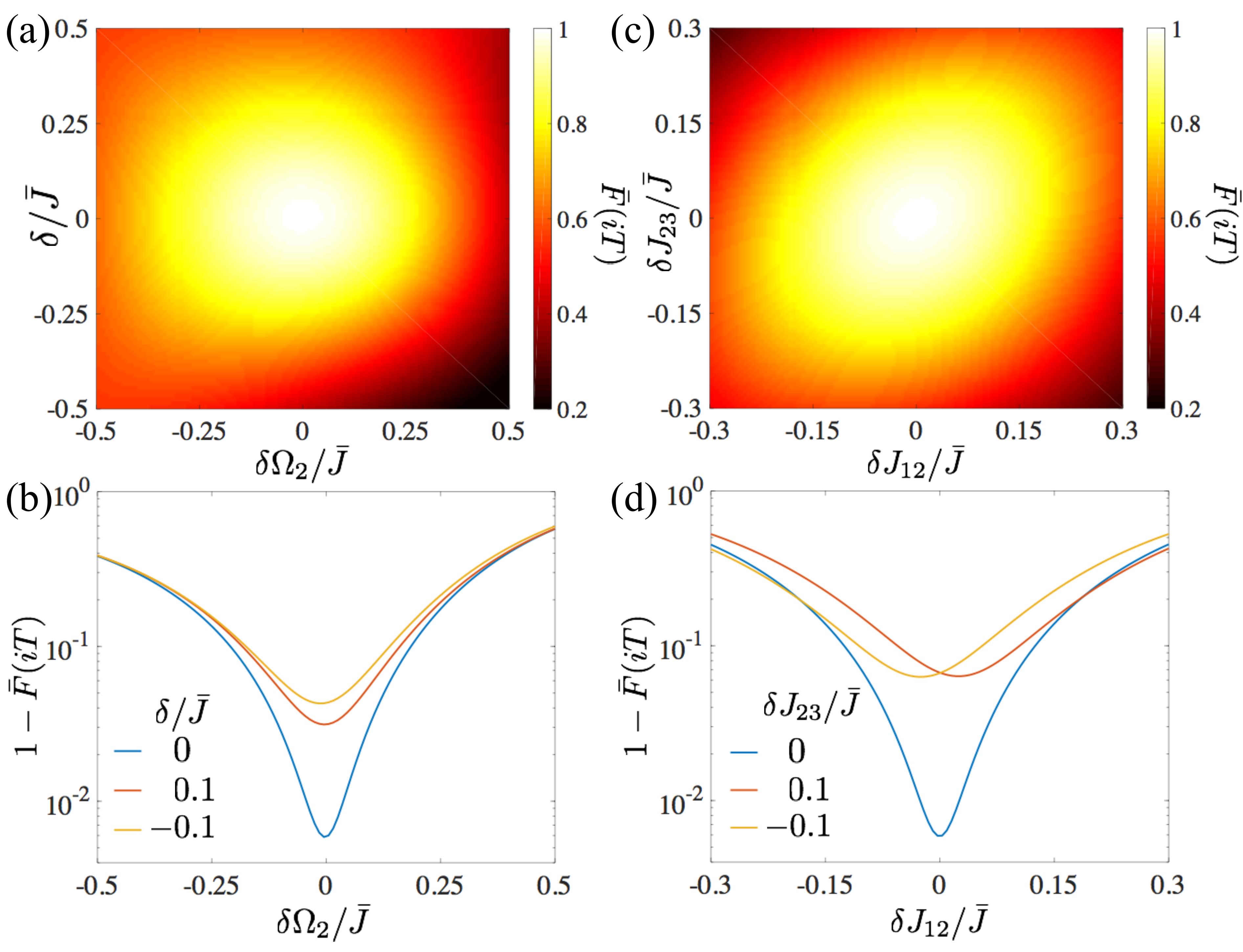}
\caption{ (a--b) Effect of detuning and Rabi frequency errors on the average fidelity of the native $i$-Toffoli gate $\bar{F}(iT)$.  (c--d) Effect of errors in the exchange coupling on $\bar{F}(iT)$.  We took $(m,n_1,n_2,n_3) =(0,1,3,0)$.  In a typical experiment $J_{ij}/2\pi =(10-20)~$MHz.}
\label{fig:staticnoise}
\end{center}
\end{figure}

At the end of the gate, there is an effective $Z$ rotation of qubit 2 (relative to the choice of  rotating frame)  that depends on the state of the other two qubits.  To remove this conditional $Z$ rotation, we arrive at the set of calibration and timing conditions
\begin{align}
\bar{J} t_{\rm dc} &= 2 \pi n_2 - \sqrt{16 n_1^2 - 3( 2m+1)^2 } \pi,  \\
\delta t_{\rm dc}& = \frac{ \Delta_2 t_{\rm dc} + 2\pi n_3}{ \Delta_2 + \bar{J}},
\end{align}
where $n_2$ is a positive integer, $n_3$ is an even integer, $t_{\rm dc} = t_{\rm dc}^1+ t_{\rm dc}^2$ and $\delta t_{\rm dc} = t_{\rm dc}^2 - t_{\rm dc}^1$.  A final set of conditions are that if $n_1$ is odd, then $n_2$ should be odd and $\Omega_2 >0$, while if $n_1$ is even, then $n_2$ should be even and $\Omega_2<0$.  The total gate time is
\be
\begin{split}
T_{\rm tot} = t_{\rm dc}+t_{\rm ac}=  \frac{\pi}{\bar{J}} &\Big[ 2 n_2 + \sqrt{4 n_1^2 -(2m+1)^2}  \\
&- \sqrt{16 n_1^2 - 3(2m+1)^2} \Big].
\label{eqn:Ttot}
\end{split}
\ee 

Under ideal conditions, the unitary $U$ takes the form
\be
U = e^{ - i \sum_i \Delta_i (t_{\rm dc} + t_{\rm ac}) s_i^z - i \bar{J}  t_{\rm ac} s_2^z }\, iT,
\ee
which is equal to the $i$-Toffoli gate up to single-qubit $Z$ rotations.
These single-qubit $Z$ rotations can be measured and then corrected in software by modifying the phase of subsequent EDSR drives on the qubits.  After this calibration step,  Fig.~\ref{fig:pop} shows that $\bar{F}(iT)$ can exceed 99$\%$ even for the maximal allowed ratio  $|\Omega_2(0,1)|/\bar{J}$, where there is significant population dynamics in the ($s_1,s_3)\ne(1,1)$ subspaces away from the $\pi$-pulse times.  

High-fidelity implementation of the i-Toffoli gate also requires careful calibration of the EDSR drive frequency and amplitude, as well as the exchange couplings.  In Fig.~\ref{fig:staticnoise}, we compute $\bar{F}(iT)$ as a function of calibration errors in the control parameters.  Here we have defined $\delta \Omega_2 = \Omega_2 - \Omega_2(m,n_1) $, $\delta = \delta_2 - \Delta_2 - \bar{J} $, and $\delta J_{ij} = J_{ij} - \bar{J} $.  In Figs.~\ref{fig:staticnoise}(a)-(b), we show the dependence of $\bar{F}(iT)$ on miscalibrations in $\delta$ and $\delta \Omega_2$, while Figs.~\ref{fig:staticnoise}(c)-(d) show the effects of miscalibrations in the exchange interactions $\delta J_{ij}$. Because the ideal implementation of the gate is a local minimum in the infidelity, for small calibration errors, $1 -\bar{F}(iT)$ has a quadratic dependence on these parameters. From Figs.~\ref{fig:staticnoise}(b) and (d), we can see that high-fidelity operation ($\bar{F}(iT)$ $>$ 90\%) is achievable for control parameter errors that are within 5\% of the average exchange coupling $\bar{J}$. 
Taking $(m,n_1,n_2,n_3) = (0,1,3,0)$ and $\bar{J}/2\pi = 20$~MHz, we find a gate operation time from Eq.~(\ref{eqn:Ttot}) of $T_{\rm tot}$ = 103 ns. In the next section, we analyze the performance of the i-Toffoli gate in the presence of $1/f$ charge noise.

\section{Sensitivity to Charge Noise}
\label{sec:noise}

In this section, we present a detailed analysis of the robustness of the $i$-Toffoli gate to time-dependent noise in the control parameters.   Spin relaxation rates in these systems are relatively slow when the Zeeman splitting is much less than the splitting to the next valley-orbital state \cite{Huang14}.  At low magnetic fields ($B^{\rm ext}$ $<$ 1 T), and even in the presence of inhomogeneous magnetic fields, spin relaxation times $T_1$ $>$ 60 ms are feasible \cite{Borjans18}.  Given that gate operations are $\sim 100$~ns, we are justified in neglecting spin relaxation processes in determining the fidelities of few-qubit gates. Furthermore, for devices based on isotopically enriched $^{28}$Si, random magnetic fields due to nuclear spins are strongly suppressed leading to increased $T_2^*$ times on the order of 10~$\mu$s \cite{Yoneda18,Sigillito19}, which is much greater than our $i$‐Toffoli gate time, but still only an order of magnitude greater than what is observed in natural Si. Thus, the dominant source of noise for $^{28}$Si devices with external magnetic field gradients likely arises from electric field noise (charge noise) that leads to time-dependent fluctuations in the parameters of the Hamiltonian, which we account for below.

To account for the charge noise, we use the parameterizations
\begin{align}
\Delta_i(t) &= \Delta_i^0 + \Delta_i^r(\bm{\Omega}) + \Delta^n_{i} v_{i}(t) ,  \\
\Omega_{i}(t) & = \Omega_{i}^0[1  + \delta  \Omega^n_{i} v_{i}(t)] ,\\
J_{ij}(t) & = J_{ij}^0 \{1 + \delta J_{ij}^r( \bm{\Omega})+ \delta J_{ij}^n  [v_i(t) + v_j(t)] \},
\end{align}
where all of these parameters are assumed to have an implicit dependence  on $(\bm{V}_{Pi},\bm{V}_{Bi})$, $v_i(t)$ is the local noise term,  $\Delta_i^0$ is the bare detuning of qubit $i$ from the EDSR reference frequency $\omega_{0i}$, $\Omega_i^0$ is the bare EDSR Rabi frequency,  and $J_{ij}^0$ is the bare exchange.    $\Delta_i^n$ is a noise sensitivity parameter that measures the change in the qubit frequency in response to the noise perturbation, $\delta \Omega_i^n$ is a noise sensitivity parameter that measures the fractional change in the EDSR Rabi frequency in response to the noise, and, similarly, $\delta J_{ij}^n$ measures the fractional change in the exchange interaction between qubits $i$ and $j$ in response to the noise, under the simplifying assumption that the exchange couples with equal magnitude to the noise field for each dot.  The parameters  $\Delta_i^r$ and $\delta J_{ij}^r$ account for additional static shifts in the detuning and exchange, respectively, that arise in the presence of the EDSR drives.

We neglect spatial correlations in the noise and take a regularized $1/f$ spectral density \cite{Ithier05,Paladino14} 
\begin{align} \label{eqn:Somega}
S(f)&=  A/f,~f_{ \ell} < f < f_c,
\end{align}
 where  $A $ is the amplitude of the $1/f$ noise, $f$ is in units of Hz, $ f_{ \ell} = (2\pi T_{\rm cal})^{-1}$ is a low-frequency cutoff, which is set by an experimental calibration time $T_{\rm cal}$, and $f_c$ is a high-frequency cutoff.  We take a white noise spectrum below $f_\ell$ and a $1/f^2$ dependence above $f_c$.  

Direct measurements of EDSR Rabi rotations can be used to determine  $\Omega_i^0$ and $\Delta_i^0$.  Ramsey or spin-echo interferometry of a single-spin with its neighboring spin in an up or down state can be used to measure $J_{ij}^0$.
For the regularized $1/f$ spectrum  in Eq.~(\ref{eqn:Somega}), the noise sensitivity parameters  are determined by the value of $T_2^*$ and the  envelope decay rates  of  the Rabi $\gamma_{r i}$ and exchange $\gamma_{e ij}$ oscillations  \cite{Ithier05}
\begin{align}
\Delta_i^n & = \sqrt{\frac{1}{A \log \big(\frac{f_c}{f_\ell} \big)}} \frac{1}{T_2^*}, \\
\delta \Omega_i^n &=  \sqrt{\frac{1}{A \log \big(\frac{f_c}{f_\ell}\big)  }}  \frac{\gamma_{ri}}{\Omega_i^0}, \\
\delta J_{ij}^n &= \sqrt{\frac{2}{ A \log \big(\frac{f_c}{f_\ell}\big) }} \frac{\gamma_{e ij}}{J_{ij}^0}.
\end{align}
The shift term $\Delta_i^r$ can be determined by simultaneously applying an off-resonant EDSR drive on the other qubits during a Ramsey sequence. To determine the static shift parameters $\delta J_{ij}^r$, far off-resonant EDSR drives can be simultaneously applied with exchange gates.

\begin{figure}[tb]
\begin{center}
\includegraphics[width=0.49\textwidth]{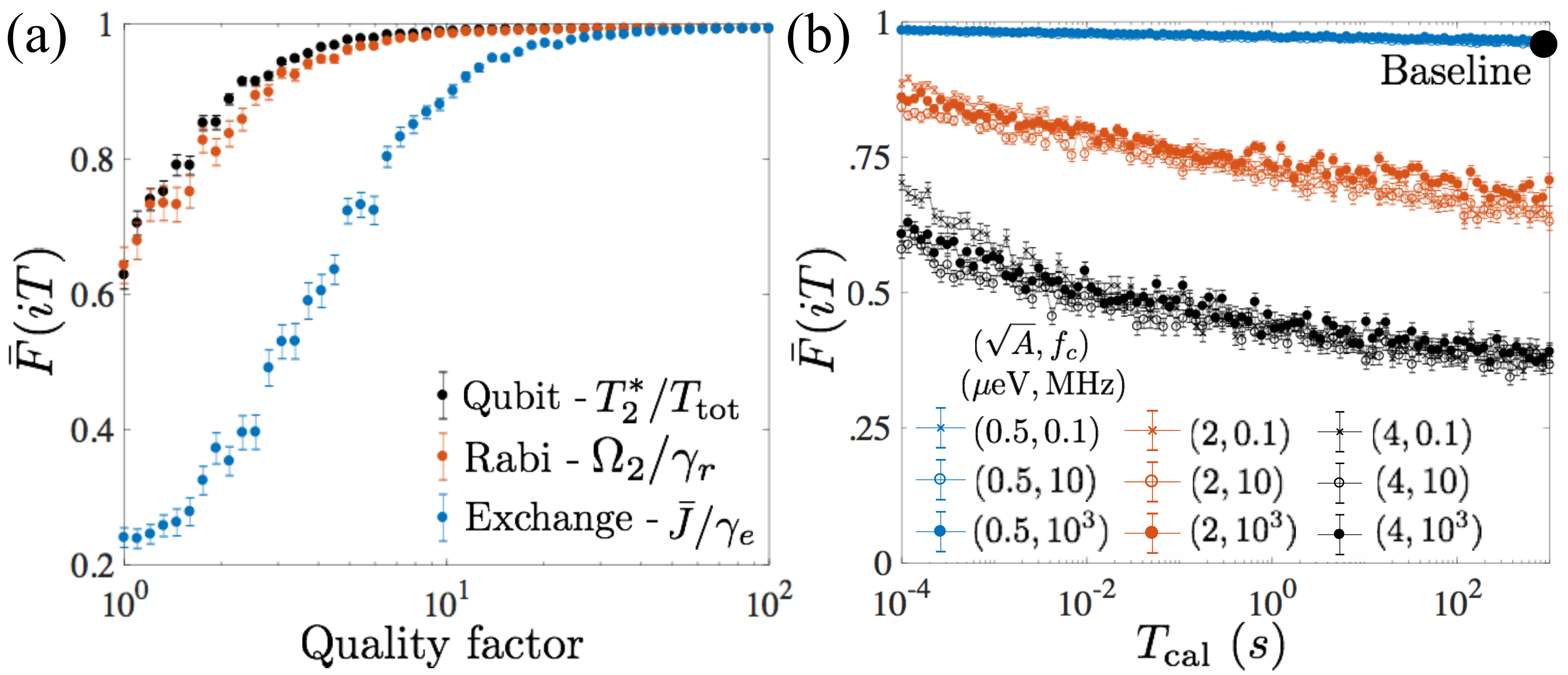}
\caption{
(a)   Dependence of $\bar{F}(iT)$ on the qubit, Rabi, and exchange quality factors of the system.  The fidelity saturates to the maximum value determined by the timing conditions for quality factors $\gtrsim 50$.  We took $\sqrt{A} = 0.5~\mu$eV, $f_c = 10~$MHz and $T_{\rm cal} = 10^3$~s.    (b) Dependence of $\bar{F}(iT)$ on $T_{\rm cal}$ for different values of $\sqrt{A}$ and $f_c$.  The noise sensitivity parameters were fixed at each value of $f_c$ at the baseline point $T_{\rm cal}^0 = 10^3$~s and $\sqrt{A}=0.5~\mu$eV \cite{Yoneda18,Mi18b}.
We used $J_{ij}/2\pi = 15~$MHz, $(m,n_1,n_2,n_3)=(0,1,3,0)$, $T_{\rm tot} = 138~$ns, and baseline values of $T_2^* = 10~\mu$s, and $\Omega_2/\gamma_r = \bar{J}/\gamma_e = 50$.   
}
\label{fig:dynnoise}
\end{center}
\end{figure}

In Fig.~\ref{fig:dynnoise}, we include the effects of time-dependent noise in a simulation of the $i$-Toffoli gate implementation. We focus on the performance of the $(m,n_1,n_2,n_3)=(0,1,3,0)$ gate because it has the shortest operation time and is, therefore, the most robust against noise. As shown in Fig.~\ref{fig:dynnoise}(a), the fast gate operation time implies that high fidelity operation can be achieved with rather modest quality factors ($T_2^*/T_{\rm tot}$, $\Omega_2/\gamma_2$, $\bar{J}/\gamma_e$ $<100$).  In Fig.~\ref{fig:dynnoise}(b), we analyze the dependence of the average gate fidelity on the calibration time, the high-frequency cutoff, and the amplitude of the $1/f$ noise.   Due to the logarithmic scaling of the noise sensitivity parameters on $f_\ell$, there is only a weak dependence of the average gate fidelity on $T_{\rm cal}$.  Similarly, changing the value $f_c$ across several orders of magnitude results in only minor changes in the average gate fidelity.  On the other hand,  modest reductions in the overall amplitude of electric field noise can significantly improve the gate fidelities.  

\section{Synthesized Toffoli gates}
\label{sec:syn}
 
\begin{figure}[tb]
\begin{center}
\includegraphics[width=0.49\textwidth]{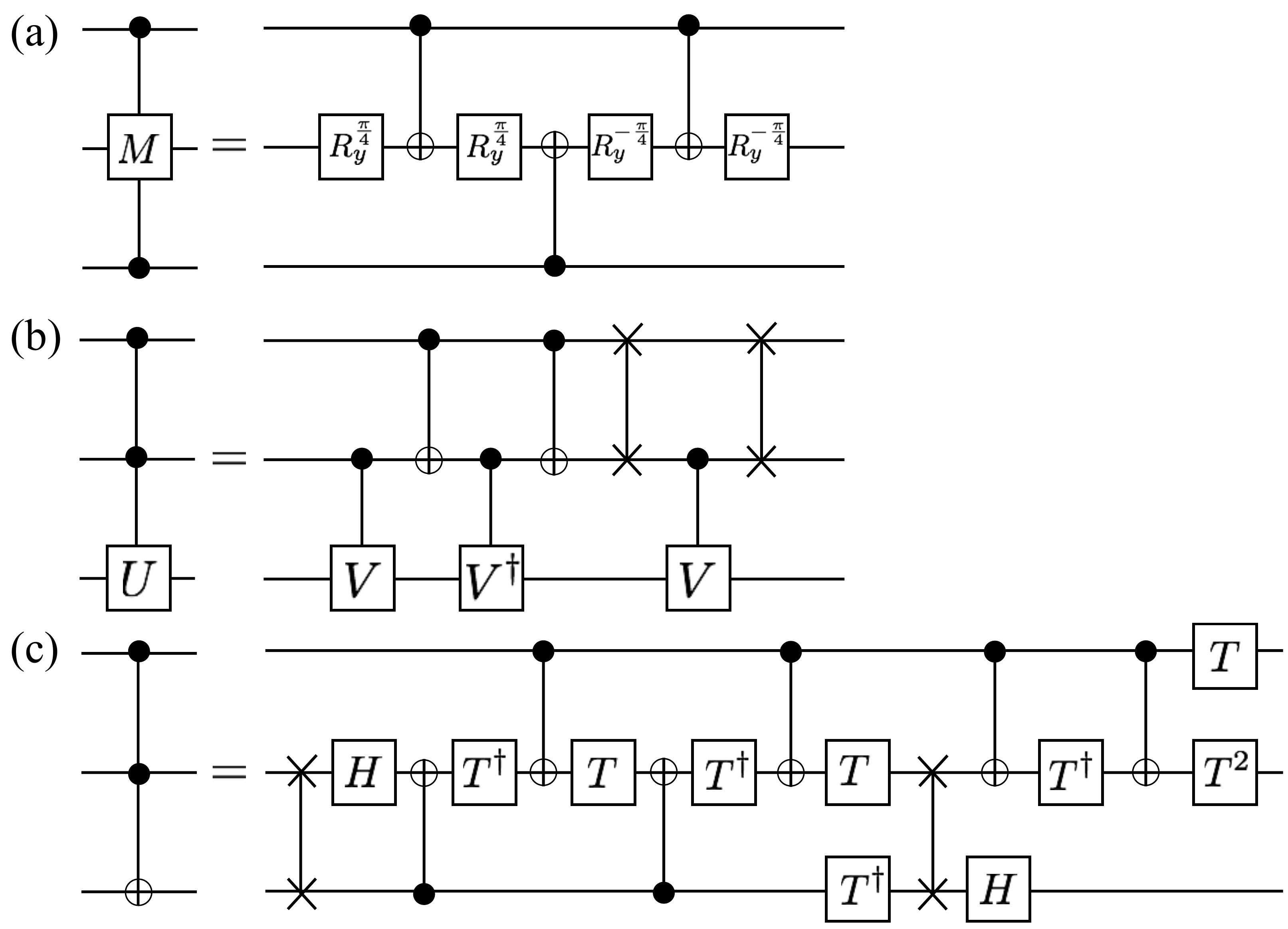}
\caption{(a) 3-qubit Margolus gate, which is equal to the Toffoli gate up to a $\pi$-phase on the state $\ket{0,1,1}$. (b) 3-qubit control-$U$ gates synthesized from SWAPs and nearest-neighbor two qubit control-$V$ gates with $V^2 = U$. For the Toffoli gate $V = e^{-i \pi/4 + i \pi \sigma_x/4}$.  (c) Synthesized Toffoli from nearest-neighbor CNOTs, SWAPs, $T$-gates,  and Hadamards ($H$).     }
\label{fig:synToff}
\end{center}
\end{figure}

In this section, we compare the performance of our Toffoli gate to Toffoli gates synthesized from two-qubit gates \cite{Barenco95,NielsenChuang}.
Efficient synthesized versions of the Toffoli gate are shown in Fig.~\ref{fig:synToff}.   The circuit in Fig.~\ref{fig:synToff}(a) realizes the Margolus gate, which is equivalent to the Toffoli gate  controlled by qubits $1$ and $3$ up to a $\pi$-phase on the state $\ket{0,1,1}$.  The Margolus gate is analogous to our $i$-Toffoli gate and, therefore, it is natural to compare the performance of these two gates.  The dominant source of errors for the Margolus gate is  likely to arise from systematic errors in the CNOT gates.  For small errors $\epsilon_{\rm CNOT}$,  this  results in an overall error rate for the Margolus gate as $\epsilon_M \approx 3 \epsilon_{\rm CNOT}$.  The operation time of the Margolus gate is 4 times longer than the $i$-Toffoli after taking into account the additional $\pi/4$ single-qubit rotations.  The $i$-Toffoli gate introduced in this work is subject to the same error mechanisms as the resonant CNOT gate studied in Ref.~\cite{Zajac18,Russ18}.  As a result, the $i$-Toffoli gate can realize up to a 3-fold reduction in error rate and 4-fold reduction in gate operation time compared to the Margolus gate.

To make a more direct comparison between our implementation and two-qubit synthesized versions of the Toffoli gate, we have to account for the ability to change the target qubit and correct the factor of $i$ in the $i$-Toffoli gate, which introduces two additional SWAP gates and one C-Phase gate.  Two efficient implementations of the Toffoli gate using nearest-neighbor two-qubit gates are shown in Figs.~\ref{fig:synToff}(b)-(c).
The circuit in Fig.~\ref{fig:synToff}(b) requires the  ability to apply controlled $\pm\pi/2$ rotations about the $x$-axis, but this is a straightforward extension of the resonantly driven CNOT gate.  This circuit involves 7 two-qubit gates, compared to the resonantly driven Toffoli that requires one $iT$ gate, and three two-qubit gates.  Thus, the error rate and operation time for the synthesized version of the Toffoli will both be as much as two times larger than our Toffoli gate implementation.  In addition, this circuit requires additional calibration of the controlled-$V$ gates.  The circuit in Fig.~\ref{fig:synToff}(c) requires the fewest number of primitive two-qubit gates and, therefore, has the least demanding calibration requirements.  However, it requires 8 two-qubit gates and 10 total rotation operations ($T$-gates can be applied in software), which will double the error rate and operation time of the gate.  

Based on this comparison, we can see that the biggest error rate reduction arises from the implementation of the $iT$ gate, which can be done in one step with our approach.  As a result, we  expect that the $iT$ gate and its generalizations to other three-qubit controlled gates will be useful options for primitive gates in compilation protocols for silicon spin-qubit quantum algorithms.

\section{Outlook and Conclusions}
\label{sec:conc}

Our implementation of the Toffoli gate naturally extends to other three-qubit controlled gates such as the Deutsch and Fredkin gates.  The Deutsch gate consists of a two-qubit controlled rotation of a third target qubit about the $x$-axis by an arbitrary angle.  Such a gate can be realized in our protocol by modifying the duration of the resonant EDSR drive on qubit 2.  The Fredkin gate is a SWAP gate on two qubits that is controlled by the state of a third qubit. To implement the Fredkin gate, one of the edge spins (e.g.\ spin 1) should be chosen as the control qubit. Due to the exchange coupling $J_{12}$, the energy of spin 2 will depend on the state of spin 1. As a result, the energy difference between states $\ket{s_1 10}$ and $\ket{s_1 01}$ will depend on $s_1$. Resonantly driving $J_{23}$ at the difference frequency of states $\ket{110}$ and $\ket{101}$ will lead to Rabi oscillations between these states that are conditioned on the orientation of spin 1. Analogous to the $i$-Toffoli gate, driving a $\pi$-pulse in this subspace leads to a direct realization of the Fredkin gate up to relative phases on each 3-qubit state in the computational basis.

To improve the fidelity and robustness of these three-qubit gates, it may be advantageous to employ dynamically corrected gates \cite{Viola09}.   Provided the noise dynamics are slow compared to the gate times, such methods can lead to substantial improvements in the gate fidelities and robustness of the gates to calibration errors \cite{Yang18,Gungordu18,CalderonVargas19}.

In conclusion, we have presented a protocol for  an efficient, high-fidelity Toffoli gate that is readily achievable in silicon spin-qubit devices based on quantum dots.  Under realistic conditions, fidelities greater than 99$\%$ are accessible with gate times at or below 100 ns.  The gate is based on a resonant EDSR drive applied to the central qubit of a 3-qubit array in the presence of finite exchange couplings $J_{12}$ and $J_{23}$.  If desired, the target qubit can be changed from the central qubit using SWAP gates.  The full implementation of the Toffoli gate is two times faster with half the error rate compared to Toffoli gates synthesized from two-qubit gates, while the $i$-Toffoli gate has a  4-fold and 3-fold reduction in time and error rate, respectively,  compared to similar 3-qubit gates.  We anticipate that the $i$-Toffoli gate, and its extensions to other three-qubit controlled gates, will be a useful primitive gate in quantum compilation approaches for silicon spin qubits.  In the near term, we expect our analysis will help guide implementations of quantum algorithms with three or more silicon spin qubits.

\begin{acknowledgements}
We thank A.\ J.\ Sigillito and D.\ M.\ Zajac for discussions. Funded by Army Research Office grant No.\ W911NF-15-1-0149, DARPA grant No. D18AC0025 and the Gordon and Betty Moore Foundation's EPiQS Initiative through Grant GBMF4535.
\end{acknowledgements}

\bibliography{QuantumDot_v3}

\end{document}